# Detecting neural activity and connectivity by perfusion-based fMRI

Jiancheng Zhuang

Dornsife Imaging Center,

University of southern California

**Abstract**

This study proposes an approach to estimate the functional localization and connectivity from CBF and BOLD signals simultaneously measured by ASL (arterial spin labeling) MRI, especially using exploratory Structural Equation Modeling analysis. In a visual task experiment, the primary visual cortices were located by analyzing the perfusion data. In the resting state experiment, two structural equation models were estimated at each voxel regarding to the sensory-motor network and default-mode network. The resulting connectivity maps indicate that supplementary motor area has significant connections to left/right primary motor areas, and inferior parietal lobules link significantly with posterior cingulate cortex and medial prefrontal cortex. The model fitting results imply that BOLD signal is more directly linked to the underlying cause of functional connectivity than CBF signal.

**Introduction**

MRI-based arterial spin labeling (ASL) techniques use water in arterial blood as a freely diffusible tracer to measure perfusion noninvasively. By labeling inflowing blood water proximal to the imaging location, the perfusion signal is subsequently obtained by subtracting the label image from the control image. Potentially, these methods can be used to map brain function and to clarify functional MRI mechanisms. It is possible that ASL contrast will provide better localization of the sites of neuronal activity than blood oxygenation level dependent (BOLD) contrast (Chuang et al., 2008).

Recently, more attention of neuroscientists has been paid to functional connectivity between cortical areas detected from resting-state fMRI data. Since ASL (arterial spin labeling) MRI technique has provided more direct observation of cerebral blood flow (CBF) fluctuations than regular BOLD-based fMRI, the perfusion MRI may reveal further information about the underlying mechanism of functional localization and connectivity (Chuang et al., 2008; Wu et al., 2009). However, there are few existing reports on detecting functional connectivity or elucidating the cause of connection networks based on perfusion MRI.

On the other hand, conventional correlation analysis across multiple spatial regions cannot reveal causal directions and the connection network between more than two regions. The Structural Equation Modeling (SEM) is an effective method to ascertain path directions and coefficients from the covariance structure in fMRI data (McIntosh and Gonzalez-Lima, 1994; Büchel and Friston, 1997; Goncalves and Hall, 2003; Penny et al., 2004; Smith et al., 2006; Zhuang et al., 2005 and 2013). Here we present an approach to estimate the functional localization and connectivity from BOLD and CBF signals simultaneously measured by ASL MRI using an exploratory SEM analysis.

**Methods and Materials**

*MRI Scan*

The ASL scans were performed on three healthy and right-handed subjects. All of them have normal or corrected vision. They were scanned on a Siemens 3T Trio/Tim system using a PASL sequence provided by Siemens company. Acquisition parameters were field of view (FOV) = 224 mm, matrix = 64 ×64, repetition time (TR) = 4 sec, echo time (TE) = 52 ms, and flip angle = 90°. Nineteen axial slices (5 mm thick without gap) were acquired from inferior to superior in a sequential order. In the task-fMRI experiment, each PASL scan with 61 acquisitions took about 4 min and it was repeated twice. In the resting state fMRI experiment, a PASL scan with 75 acquisitions took about 5 minutes.

*fMRI Experimental Design*

In a task-related fMRI scan, it is a block design study with two alternating conditions, the left and right checkerboard conditions. In each condition, a black and white checkerboard was presented at the left or the right visual field of the subject for 12s. Each condition was repeated for 10 times. In the resting-state experiment, the subjects were instructed to close eyes but stay awake, and not to perform any mental task.

*Statistical Analysis*

Perfusion fMRI data were firstly processed on-line by the perfusion module and GLM module accompanying with this pulse sequence. We set the GLM parameters matching with this fMRI design. The data were also analyzed with SPM software (Wellcome Department of Imaging Neuroscience, London, UK). The perfusion-weighted were realigned to correct head motion and coregistered to the anatomical image of individual subject. The perfusion images were smoothed with an 8mm Full-Width Half-Maximum (FWHM) Gaussian kernel. The pair-wise subtraction between labeled and control images was performed to obtain CBF weighted images. Thereafter, the data were high-pass filtered at 0.08 Hz to isolate the CBF signal. The labeled and control images of the low-pass filtered (<0.07 Hz) PASL data were summed together to generate BOLD signal (Chuang et al., 2008). General linear model was used to model the changes of perfusion elicited by the two conditions while no temporal filtering or smoothing was applied. Statistical parametric maps for each subject were generated at an uncorrected threshold of p<0.005 with an extent threshold of 10 voxels and overlayed onto the anatomical image of each subject.

In SEM analysis of the resting state data, the left and right primary motor areas (L/R M1) were chosen as two ROIs for the detection of sensory-motor network, and the posterior cingulate cortex (PCC) and medial prefrontal cortex (MPFC) were selected as two ROIs for the detection of default-mode network. The signal of each voxel within the brain was regarded as the third observation in the two structural equation models. Each model consists of two connections start from the unknown region to the two predefined ROIs (Figure 3). For every voxel, the models were estimated by statistical indices such as Adjusted Goodness-of-Fit Index (AGFI) with a significant t-value (>1.96) on each path (Equations 1 and 2).

$$GFI = 1-tr[(S^{-1}S-I)^2]/tr[(S^{-1}S)^2] \qquad (1)$$

where tr indicates the trace operation, S is the covariance matrix of signals, and S is the estimated S;

$$AGFI = 1-[p(p+1)/2df](1-GFI) \qquad (2)$$

where p the number of observations, and df the degree of freedom in the model.

The AGFI at each voxel was saved and displayed in the final connectivity map. The model fit indices were compared between connectivity maps obtained from CBF and BOLD signals.

**Results and Discussions**

The real-time analysis results in the activation map, which shows the left primary visual cortex was activated during the right visual field checkerboard condition, while the right primary visual cortex was activated during the left visual field checkerboard condition, as what we expected (Figure 1).

From the off-line analysis, the left primary visual cortex was also found activated during the right visual field checkerboard condition and the right primary visual cortex was activated during the left visual field checkerboard condition in both subjects (Figure 2). It is consistent with our on-line results (Figure 1).

**Figure 1**. The on-line results showing the activity map from the visual task experiment

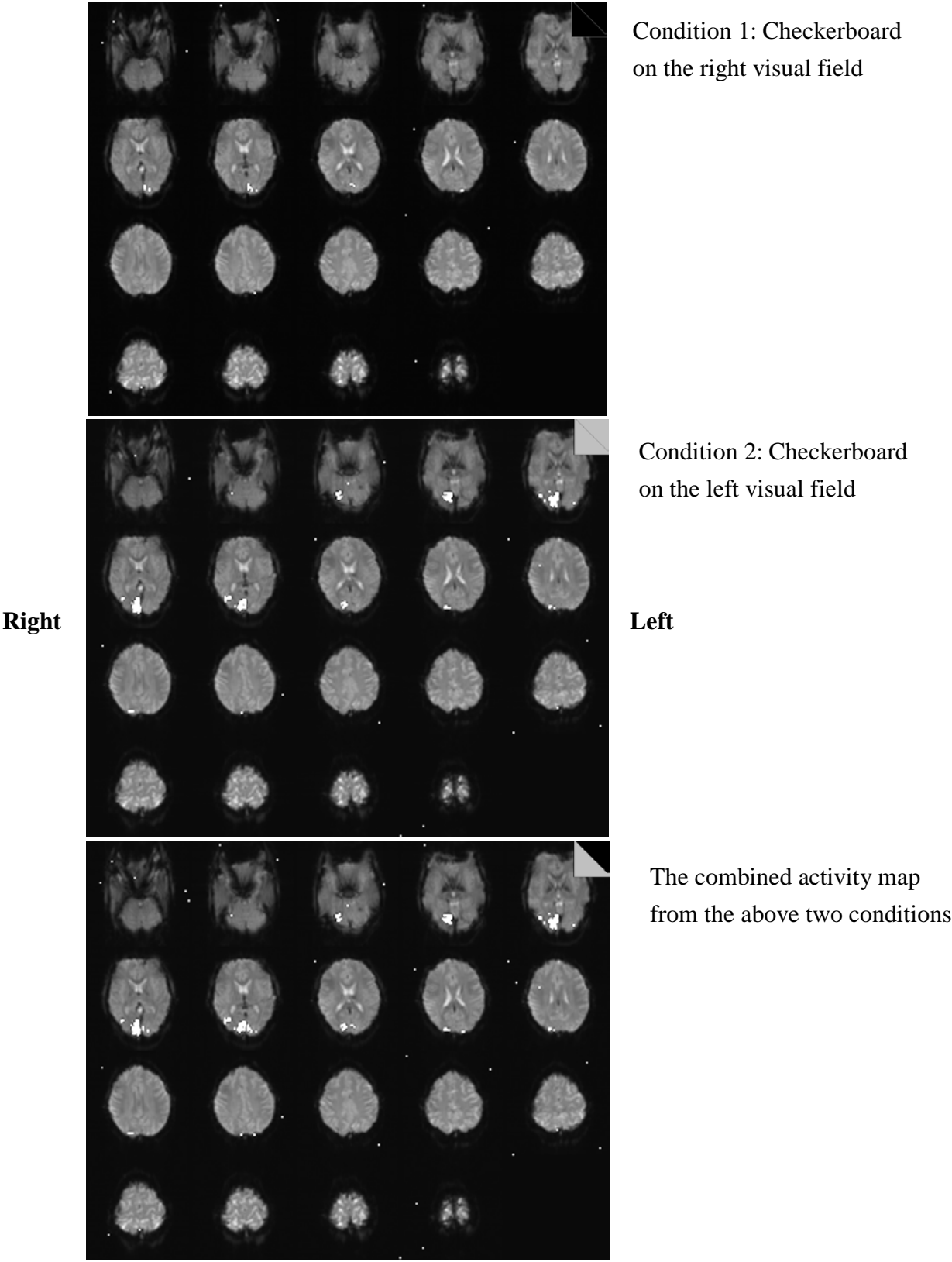

Condition 1: Checkerboard on the right visual field

Condition 2: Checkerboard on the left visual field

**Right** **Left**

The combined activity map from the above two conditions

**Figure 2**. The off-line analyzed results showing the activity map from the visual task experiment

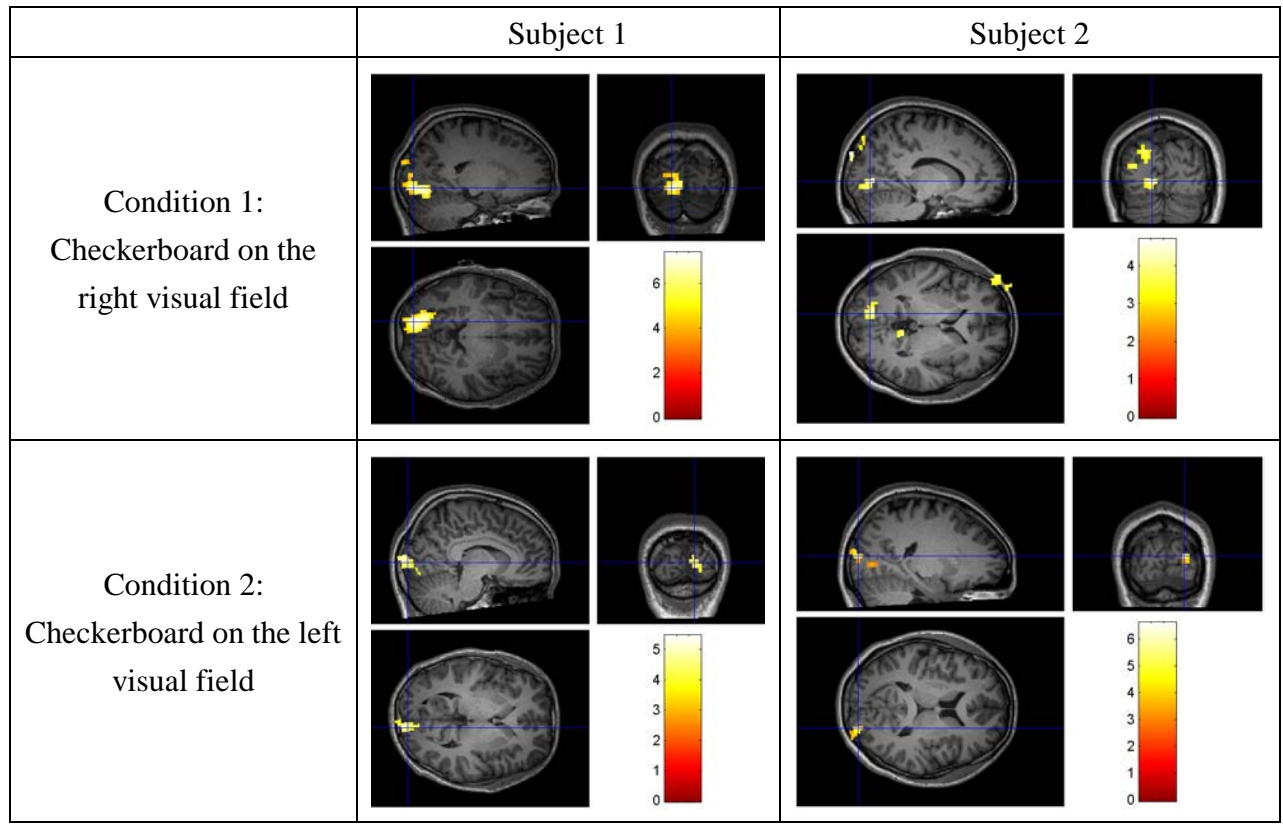

The representative connectivity maps resulted from SEM are illustrated in the Figure 3. The connectivity maps were found to be reproducible across subjects, and were similar between CBF and BOLD results. It indicates that the supplementary motor area (SMA) has significant connections to both LM1 and RM1 in the sensory-motor network, and the bilateral inferior parietal lobules (IPL) link significantly with PCC and MPFC in the default-mode network, which is consistent with the neuroanatomical evidence and existing results from functional connectivity researches using BOLD-based fMRI. Parts of predefined ROIs shown in the connectivity map can be interpreted as direct interactions between the two ROIs, which is also often found in the fMRI studies of functional connectivity. The largest standardized residual (the latent variable) obtained from SEM fitting at CBF signal is much larger than that obtained at BOLD signal (Table 1), which implies BOLD signal is more directly linked to the neuronal cause of functional connectivity.

**Figure 3.** The structural equation models and resulting connectivity maps from resting-state data

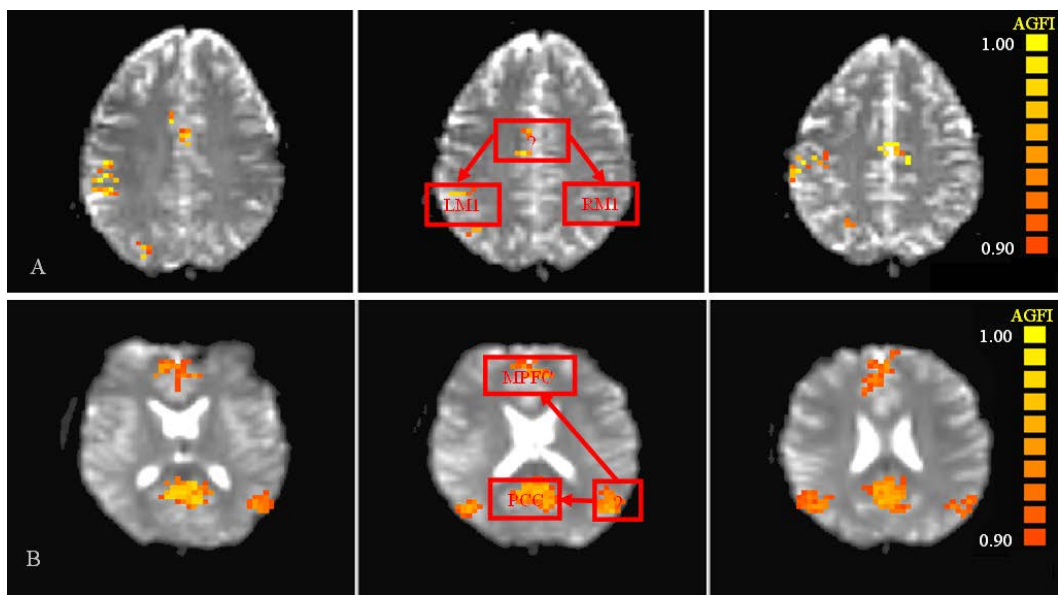

The structural equation models (middle in red), and resulting connectivity maps in the sensory-motor network (A) and default-mode network (B) detected from CBF signals, which are similar to the maps obtained from BOLD signals (Zhuang et al., 2013).

**Table 1.** Largest standardized residual averaged (and its standard deviation) from SEM fittings at each network and each signal from resting-state data

|  | **CBF** | **BOLD** |
|---|---|---|
| **Sensory-motor Network** | 0.0417 | 0.0176 |
|  | (0.0133) | (0.0091) |
| **Default-mode Network** | 0.0923 | 0.0285 |
|  | (0.0165) | (0.0102) |

The largest standardized residual obtained from SEM fitting at CBF signal is much larger than that obtained at BOLD signal, which implies BOLD signal has better SNR or has less latent variables linked to the neuronal cause of functional connectivity.

The present approach of SEM analysis on the perfusion MRI data allows us to search the possible functional connections at each brain region directly from CBF signals. When several areas are involved in a functional connection network, which is often the case, such connections become too complicated to be ascertained simply via correlation analysis. SEM produced maps and estimations circumvent this difficulty, and can be used to further examine complicate network models and possible physiological mechanisms underlying functional connectivity results based on BOLD or perfusion MRI.

In conclusion, the perfusion-based fMRI can detect the neural activity and connectivity accurately. More detailed comparison between conventional fMRI and this technique is still in progress.

**Reference**

Büchel, C., Friston, K.J., 1997. Modulation of connectivity in visual pathways by attention: cortical interactions evaluated with structural equation modelling and fMRI. Cereb. Cortex 7, 768–778.

Chuang, K.H., van Geldern, P., Merkle, H., Bodurka, J., Ikonomidou, V.N.., Koresky, A.P., Duyn, J.H., Talagald. S.L., 2008. Mapping resting-state functional connectivity using perfusion MRI. Neuroimage 40, 1595-1605.

Goncalves, M.S., Hall, D.A., 2003. Connectivity analysis with structural equation modelling: an example of the effects of voxel selection. NeuroImage 20, 1455–1467.

McIntosh, A.R., Gonzalez-Lima, F., 1994. Structural equation modeling and its application to network analysis in functional brain imaging. Hum. Brain Mapp. 2, 2–22.

Penny, W.D., Stephan, K.E., Mechelli, A., Friston, K.J., 2004. Modelling functional integration: a comparison of structural equation and dynamic causal models. NeuroImage 23, S264–S274.

Smith, J.F., Chen, K., Johnson, S., Morrone-Strupinsky, J., Reiman, E.M., Eric, M., Nelson, A., Moeller, J.R., Alexander, G.E., 2006. Network analysis of single-subject fMRI during a finger opposition task. NeuroImage 32, 325–332.

Wu, C.W., Gu, H., Lu, H., Stein, E.A., Chen, J-H., Yang, Y., 2009. Mapping functional connectivity based on synchronized $CMRO_2$ fluctuations during the resting state Neuroimage 45, 694-701.

Zhuang, J., LaConte, S., Peltier, S., Zhang, K., Hu, X., 2005. Connectivity exploration with structural equation modeling: an fMRI study of bimanual motor coordination. NeuroImage 25, 462–470.

Zhuang, J., Peltier, S., He. S., LaConte, S., Hu, X., 2008. Mapping the connectivity with structural equation modeling in an fMRI study of shape-from-motion task. NeuroImage 42, 799–806.